\documentstyle[prb,aps]{revtex}

\begin{document}
\twocolumn
\draft

\title{Relativistic separable dual-space Gaussian Pseudopotentials\\
 from H to Rn}

\author{ C.~Hartwigsen, S.~Goedecker, J.~Hutter\\
Max-Planck-Institut f\"ur Festk\"orperforschung, Stuttgart, Germany}

\maketitle

\begin{abstract}
We generalize the concept of separable dual-space Gaussian
pseudopotentials to the relativistic case. This allows us to construct
this type of pseudopotential for the whole periodic table and we
present a complete table of pseudopotential parameters for all the
elements from H to Rn. The relativistic version of this
pseudopotential retains all the advantages of its nonrelativistic
version.  It is separable by construction, it is optimal for
integration on a real space grid, it is highly accurate and due to its
analytic form it can be specified by a very small number of
parameters. The accuracy of the pseudopotential
is illustrated by an extensive series of molecular calculations.
\end{abstract}

\pacs{71.15.Hx}

\section{Introduction}
Pseudopotentials are a well established tool in ab initio structure
calculations of molecules and solids. First, by replacing the atom by
a pseudo atom the number of orbitals which have to be calculated is
reduced and second, the size of the basis set can substantially be
reduced because the pseudo wave functions are smoother than their
all-electron counterparts. In addition relativistic effects which are
relevant for heavier elements can be included in the pseudopotential
construction so that a non-relativistic calculation can reproduce
these

In 1982 Bachelet, Hamann and Schl\"uter \cite{a50} published a list
of pseudopotentials for all elements up to Pu, that has found
widespread application. There have been many attempts to improve the
pseudopotential transferability and their numerical efficiency
since. One major advance was the introduction of a separable form by
Kleinmann and Bylander \cite{a51}, that significantly reduces the
computational effort for the calculation of the nonlocal part
especially when using a plane wave basis set. Gonze, Stumpf and
Scheffler\cite{SGS1} investigated the Kleinmann-Bylander form
carefully and published a list\cite{SGS2} of pseudopotentials for many
elements up to Xe. Goedecker {\sl et al.}\cite{x1} proposed a new
dual-space Gaussian type pseudopotential which is separable and
satisfies an optimality criterion for the real space integration of
the nonlocal part. For large systems there is only a quadratic scaling
with respect to the system size if the integration of the nonlocal
part is performed on a real space grid compared to a cubic scaling if
a Fourier space integration is used \cite{goed3}. In contrast to most
other pseudopotential construction methods Goedecker {\sl et al.} also
included unoccupied orbitals in their method thereby generating highly
transferable pseudopotentials. Goedecker {\sl et al.} gave the non
relativistic pseudopotential parameters for the first two rows of the
periodic system and showed that their pseudopotentials give highly
accurate results in molecular calculations.  They obtained results
which are much closer to the quasi-exact all-electron
LDA~\cite{lda1,lda2} (local density approximation) value than what is
obtained in all-electron calculations with a standard Gaussian 6-31G$^*$ basis sets. With other
words the errors due to the pseudopotential approximation were much
smaller than the errors in a all-electron calculation introduced by
incomplete basis sets.

In this paper we give the parameters of dual-space Gaussian
pseudopotentials for all elements from H to Rn. In contrast to
Goedecker {\sl et al.} all pseudopotentials are now generated on the
basis of a fully relativistic all-electron calculation, i.e. by
solving the two component Dirac equation. The generalization of the
norm-conservation property to the relativistic case proposed by
Bachelet and Schl\"{u}ter \cite{a543} is used for the construction. We
also introduced some slight modifications of the analytic form of the
pseudopotential.  The parameters are given in the context of the local
density approximation. Even though the parameters
change only slightly if the pseudopotential is constructed 
within the framework of a generalized gradient
approximation~\cite{PBE,BLYP} (GGA) functional, 
we found that molecular properties are less accurately
described if LDA pseudopotentials are inserted in 
a molecular calculation using GGA's.
Since it is not possible to construct pseudopotentials
tables for all current GGA schemes, a program that can construct
pseudopotentials for the most common GGAs can be obtained from the
authors.

\section{Form of the Pseudopotential }\label{fotp}
The local part of the pseudopotential is given by
\begin{eqnarray}\label{eq1}
V_{\rm loc}(r)&=&\frac{-Z_{\rm ion}}{r}
\mbox{erf}\left(\frac{r}{\sqrt{2}{r_{\rm loc}}}\right) 
+ \exp\left[-\frac{1}{2}\left(\frac{r}{{r_{\rm loc}}}\right)^2\right]
\nonumber\\
&&\times \left[ C_1 
             +C_2 \left(\frac{r}{{r_{\rm loc}}}\right)^2
             +C_3 \left(\frac{r}{{r_{\rm loc}}}\right)^4
             +C_4 \left(\frac{r}{{r_{\rm loc}}}\right)^6
       \right]
\end{eqnarray}
where erf denotes the error function. $Z_{\rm ion}$ is the ionic
charge of the atomic core, i.e. the total charge minus the charge of
the valence electrons.  The nonlocal contribution $V_l({\bf
r},{\bf r}')$ to the pseudopotential is a sum of separable terms
\begin{equation}\label{eq2}
 V_l({\bf r},{\bf r}')= \sum_{i=1}^3 \sum_{j=1}^3
\sum_{m=-l}^{+l} Y_{l,m}(\hat{\bf r})\; p_i^l(r)\;  h_{i,j}^l\;
p_j^l(r')\;  Y^*_{l,m}(\hat{\bf r}')
\end{equation}
where $Y_{l,m}$ are the spherical harmonics and $l$ the angular
momentum quantum number.  The projectors $p_i^l(r)$ are Gaussians of
the form
\begin{equation}
p_i^l(r)=\frac{\sqrt{2} r^{l+2(i-1)}\exp\left(-\frac{r^2}{2{r_l}^2}\right)}
{{r_l}^{l+(4i-1)/2} \sqrt{\Gamma(l+\frac{4i-1}{2}) }}
\end{equation} 
where $\Gamma$ denotes the gamma function.  The projectors satisfy the
normalisation condition
\begin{equation}\label{eq5}
\int_0^\infty p_i^l(r) p_i^l(r)r^2dr=1 \; .
\end{equation}

It is a special property of our pseudopotential that it has also an
analytical form if expressed in reciprocal space.  The Fourier
transform of the pseudopotential is given by
\begin{eqnarray}
V_{\rm loc}(g)&=&-\frac{4\pi Z_{\rm ion}}{\Omega g^2}
                 e^{-\frac{(g {r_{\rm loc}})^2}{2}} \nonumber \\
&&  +\sqrt{8\pi^3}\frac{{r_{\rm loc}}^3}{\Omega} e^{-\frac{(g {r_{\rm loc}})^2}{2}}
    \times \left\{C_1+C_2\left(3-g^2{r_{\rm loc}}^2\right) \right. \nonumber \\
&&  +C_3\left(15-10(g r_{\rm loc})^2 + (g r_{\rm loc})^4\right) +C_4\left(105\right.  \nonumber \\
&& \left. \left. -105(g r_{\rm loc})^2 +21 (g r_{\rm loc})^4 - (g r_{\rm loc})^6\right)
                 \right\}
\end{eqnarray}
for the local part and
\begin{eqnarray}
 V_l({\bf g},{\bf g}') &=&  (-1)^l \sum_{i=1}^3\sum_{j=1}^3
\sum_{m=-l}^{+l}
Y_{l,m}(\hat{\bf g})\;
p_{i}^l(g)\nonumber\\
&&\times  h_{i,j}^l\; p_{j}^l(g')\; Y^*_{l,m}(\hat{\bf g}')
\end{eqnarray}
for the nonlocal part. The Fourier transform of the projectors
$p^l_i(r)$ can be calculated analytically and for the relevant cases
one obtains
\begin{eqnarray}
p_1^{l=0}(g)&=&\frac{4 \sqrt{2{r_0}^3} \pi^{5/4}}
                  {\sqrt{\Omega} \exp[\frac{1}{2}(g {r_0})^2]} \; ,\\
p_2^{l=0}(g)&=&\frac{8 \sqrt{\frac{2{r_0}^3}{15}} \pi^{5/4} (3-g^2{r_0}^2)}
                  {\sqrt{\Omega} \exp[\frac{1}{2}(g {r_0})^2]} \; ,\\
p_3^{l=0}(g)&=&\frac{16 \sqrt{\frac{2{r_0}^3}{105}} \pi^{5/4} (15-10g^2{r_0}^2+g^4{r_0}^4)}
                  {3 \sqrt{\Omega} \exp[\frac{1}{2}(g {r_0})^2]} \; ,\\
p_1^{l=1}(g)&=&\frac{ 8 \sqrt{\frac{{r_1}^5}{3}} \pi^{5/4} g }
                  {\sqrt{\Omega} \exp[\frac{1}{2}(g {r_1})^2]} \; ,\\
p_2^{l=1}(g)&=&\frac{ 16 \sqrt{\frac{{r_1}^5}{105}} \pi^{5/4} g (5-g^2{r_1}^2)}
                  {\sqrt{\Omega} \exp[\frac{1}{2}(g {r_1})^2]} \; ,\\
p_3^{l=1}(g)&=&\frac{ 32 \sqrt{\frac{{r_1}^5}{1155}} \pi^{5/4} g (35-14g^2{r_1}^2+g^4{r_1}^4)}
                  {3 \sqrt{\Omega} \exp[\frac{1}{2}(g {r_1})^2]} \; ,\\
p_1^{l=2}(g)&=&\frac{ 8 \sqrt{\frac{2{r_2}^7}{15}} \pi^{5/4} g^2 }
                  { \sqrt{\Omega} \exp[\frac{1}{2}(g {r_2})^2]} \; ,\\
p_2^{l=2}(g)&=&\frac{ 16 \sqrt{\frac{2{r_2}^7}{105}} \pi^{5/4} g^2 (7-g^2{r_2}^2)}
                  { 3 \sqrt{\Omega} \exp[\frac{1}{2}(g {r_2})^2]} \; ,\\
%p_3^{l=2}(g)&=&\frac{ 32 \sqrt{\frac{2{r_2}^7}{15015}} \pi^{5/4} g^2 (63-18g^2{r_2}^2+g^4{r_2}^4}
%                  { 3\sqrt{\Omega} \exp[\frac{1}{2}(g {r_2})^2]} \; ,\\
p_1^{l=3}(g)&=&\frac{ 16 \sqrt{\frac{{r_3}^9}{105}} \pi^{5/4} g^3}
                  { \sqrt{\Omega} \exp[\frac{1}{2}(g {r_3})^2]} \; .
\end{eqnarray}

In both real and Fourier space the projectors have the form of a
Gaussian multiplied by a polynomial. Due to this property the
dual-space Gaussian pseudopotential is the optimal compromise between
good convergence properties in real and Fourier space. The
multiplication of the wave function with the nonlocal pseudopotential
arising from an atom can be limited to a small region around the atom
as the radial projectors $p^l_i(r)$ asymptotically tend to zero
outside the covalent radius of the atom. In addition, a very dense
integration grid is not required as the nonlocal pseudopotential is
reasonably smooth because of its good decay properties in Fourier
space.

The use of this form for the pseudopotential is also very advantageous if
atom centered basis functions are used instead of plane waves.  Because
of the separability all three-center integrals are products of two-center 
integrals and so only these two-center integrals have to be calculated. 
If atom centered Gaussian type orbitals are used, these two-center integrals 
can easily be evaluated analytically.

In the relativistic case the spin orbit coupling splits up all
orbitals with $l>0 $ into a spin up and spin down orbital with overall
angular momentum $j=l\pm 1/2$.  So for each angular-momentum $l>0$ one
spin up and spin down orbital with different wave functions and
pseudopotentials exist. Following Bachelet and Schl{\"u}ter
\cite{a543} we give a weighted average and difference potential of
these potentials. The average pseudopotential is conveniently defined
as
\begin{equation}
V_l({\bf r},{\bf r}')=\frac{1}{2l+1}\left(l
V_{l-1/2}({\bf r},{\bf r}') + (l+1) V_{l+1/2}
({\bf r},{\bf r}')\right)
\end{equation}
weighted by the different $j$ degeneracies of the $l \pm 1/2$
orbitals. The difference potential describes the spin orbit coupling 
and is defined as
\begin{equation}
\Delta V^{\rm
SO}_l({\bf r},{\bf r}')=\frac{2}{2l+1}\left(V_{l+1/2}({\bf r},{\bf r}')
- V_{l-1/2}({\bf r},{\bf r}')\right).
\end{equation}
The total pseudopotential is then given by 
\begin{equation}\label{eq20}
V({\bf r},{\bf r}') = V_{\rm loc}(r)\delta({\bf r}-{\bf
r}') + \sum_l V_l({\bf r},{\bf r}') +
\Delta V^{\rm SO}_l({\bf r},{\bf r}'){\rm \bf
{L}\cdot{S}}.
\end{equation}
where $V_{\rm loc}(r)$ and $V_l({\bf r},{\bf r}'$) are now
scalar relativistic quantities but with the same form
(eq.(\ref{eq1},\ref{eq2})) as the non relativistic case.  To express
$\Delta V^{\rm SO}_l({\bf r},{\bf r}')$ we also use
eq.(\ref{eq2}) just replacing the $h^l_{i,j}$ by different parameters
$k^l_{i,j}$, i.e.
\begin{equation}
\Delta V_l^{\rm SO}({\bf r},{\bf r}')=\sum_{i=1}^3 \sum_{j=1}^3
\sum_{m=-l}^{+l}Y_{l,m}(\hat{\bf r}) 
p_i^l(r) k_{i,j}^l p_j^l(r') Y^*_{l,m}(\hat{\bf r}') .
\end{equation} 
Neglecting the contributions from $\Delta V_l^{\rm SO}({\bf
r},{\bf r}')$ in eq.(\ref{eq20}) gives the average potential that
contains all scalar parts of the relativistic pseudopotential whereas
the total potential contains relativistic effects up to order
$\alpha^2$.

\section{Determination of the pseudopotential parameters}
The parameters of the pseudopotentials were found by minimising the
differences between the eigenvalues and the charges within an atomic
sphere of the all-electron and the pseudo atom. In most cases the the
radius of the atomic sphere was taken to be the covalent radius of the
atom.  For consistency we always performed a fully relativistic
calculation for the all-electron atom even when relativistic effects
are negligible. The exchange and correlation energy was calculated
with the functional given by Goedecker {\sl et al.}\cite{x1} This
functional reproduces very well the Perdew-Wang\cite{x2} 1992 but is
much easier to compute. To ensure transferability of the
pseudopotential we also considered the next two or three higher
unoccupied orbitals for each angular-momentum and the lowest orbital
of the next two unoccupied angular-momentum. However in our
calculations we never exceeded $l_{\rm max}$=3. The atom was put in an
external parabolic confining potential to have well defined unoccupied
orbitals.  The pseudopotential parameters given in table \ref{table1}
typically reproduce the eigenvalues of the occupied orbitals with an
error of less than 10$^{-5}$ a.u. and for the unoccupied orbitals to
within 10$^{-3}$ a.u. Pseudopotentials containing semi-core electrons
(next Sec.)  are an exception as the errors for the semi-core orbitals
are usually larger than for the valence orbitals. In many cases we
found it unnecessary to include all unoccupied orbitals in our fitting
procedure.  For most cases the inclusion of only the first unoccupied
orbital for an angular-momentum results in comparable good results
for the following higher unoccupied orbitals. Nevertheless we always
checked all-electron and pseudo eigenvalues and charges of the unoccupied
orbitals to verify this.

It has already been discussed by Goedecker {\sl et al.} that our
fitting procedure yields pseudopotentials that obey the
norm-conservation\cite{a53} condition and meet several additional
conditions\cite{a49,x3,goed9}, such as extended norm conservation and
hardness, thereby leading to pseudopotentials of a very high quality.

In table \ref{table3} we give the transferability errors for several
excited and ionized states for some elements.

The construction of our pseudopotential differs somehow from the usual
method because we fit the pseudopotential parameters that give the best
overall representation for the eigenvalues and charges of several
orbitals rather than producing pseudo wave functions that are
identical to their all-electron counterparts beyond some cutoff
radius. Therefore the wave functions of the pseudo atom and
all-electron atom approach each other only exponentially.  Nevertheless the
difference is very small beyond the core region as can been seen from
Fig.~\ref{fig1}. A second consequence of our fitting procedure is that
contrary to most other pseudopotential construction methods the local
part of our pseudopotential does not correspond to a certain wave
function.

It is as special feature of our method that we fit our parameters
directly against the all-electron eigenvalues and charges rather than
fitting analytical or numerical potentials that reproduce pseudo wave
functions which themselves are constructed from their all-electron
counterparts. Therefore our pseudopotentials require significant fewer
parameters than those tabulated by Bachelet, Hamann and Schl\"uter
\cite{a50}.  During the generation of our pseudopotentials we found that
there is in general no single minimal parameter set that gives the
best overall pseudopotential for one atom. This finding is different
to the former study of Goedecker {\sl et al.} where only the first two
rows of the periodic table have been considered. We always tried to
use the minimum parameter set which is sufficient to reproduce the
desired accuracy of the fitted eigenvalues and charges. Identical
parameter sets were used for comparable elements, i.e. the same
parameter set was used for all $3d$-elements or the $4d$-elements,
respectively.  The fitting of the pseudopotential parameters is
numerically demanding as many local minima exist so that sometimes up
to some $10^5$ pseudopotential evaluations are necessary until one
finds good parameter values. We used a slow Simplex-Downhill
algorithm\cite{b1} for the optimisation that proved to be much more
robust than more sophisticated methods.  The pseudopotential parameter
${r_{\rm loc}}$ was set by hand except for the first row because this
parameter is not easy accessible by our fitting procedure. For many
elements we generated and tested pseudopotentials with different
values of ${r_{\rm loc}}$. After selection of the optimum
pseudopotentials the ${r_{\rm loc}}$ values for the elements in
between were interpolated so that no discontinuities occur.

\section{semi-core electrons}
For many atoms there is no unambiguous separation of the electronic
system into a well isolated core and valence shell. For example it is
well known the ($n$-1) $p$-levels of the the heavy alkali atoms are
relatively shallow in energy and extended in space. The $3d$-wave
functions of the $3d$-elements are strongly localized so that there is
a significant overlap with the $3(s,p)$-wave functions although the
later are much lower in energy than the $3d$- and $4(s,p)$-valence
wave functions. The same is true for the $4d$- and
$5d$-elements. Analogous the $4f$-wave functions of the $4f$-elements
are so localized that they overlap with the $5s$- and $5p$-wave
functions.  In all these cases where a non negligible overlap between
valence and core wave functions exists the frozen-core approximation
underlying the construction of all pseudopotentials is not well
satisfied. One way to overcome this problem is the inclusion of a
nonlinear core correction \cite{a407} that considers the contribution
of the core charge to the exchange-correlation potential. The other
more straight forward solution is the explicit inclusion of the
semi-core electrons into the pseudopotential. In this work we decided
for the second method. This ensures that our semi-core
pseudopotentials still can be used with programs where non linear
corrections are not considered. In addition the explicit inclusion of
the semi-core electrons ensures that our pseudopotentials still work
well for systems where non linear core corrections fail.  It is
unnecessary that the eigenvalues and charge distribution of the
semi-core wave functions have the same accuracy as the valence wave
functions of the pseudo atom. We always tried to generate semi-core
pseudopotentials with semi-core pseudo wave functions that are as
smooth as possible but still yield accurate results for the valence
wave functions.  Therfore the error for the eigenvalues of semi-core
wave functions for our pseudopotentials is within 10$^{-3}$ to
10$^{-2}$ a.u. which is about 3 orders of magnitude worse than the
typically error for the valence wave functions.

The choice which electrons are treated as semi-core electrons also
depends on the required accuracy. As we were interested to generate
pseudopotentials that can be used together with plane wave basis sets
within a reasonable computational effort we tried to include not too
many semi-core electrons into our pseudopotentials. Our semi-core
pseudopotentials for the group Ia and IIa elements, the transition
metals of group IIIb to VIIIb and the lanthanides trade the ($n$-1)$s$-
and the ($n$-1)$p$-electrons as semi-core electrons. For the elements of
group Ib, IIb and IIIa (except B and Al) all electrons of the
completely filled $nd$-shell are treated as semi-core electrons.

For all elements mentioned above semi-core wave functions improve the
description of highly positive charged ions. In table~\ref{table3} the
transferability error of two Ti pseudopotentials is listed for several
states. For most states the calculated excitation energies are much
closer to the all-electron values for the Ti semi-core pseudopotential
including the $3s$- and $3p$-semi-core electrons. This is most significant
for the $4s^04p^03d^0$ state which corresponds to a
Ti$^{4+}$-ion. For the 4$e$-pseudopotential the error is 0.1 Hartree
but only 0.28$\times$10$^{-2}$ Hartree for the 12$e$-semi-core
pseudopotential.

Pseudopotentials with semi-core wave functions always require higher
computational effort. They contain more electrons and larger basis
sets are necessary for a sufficient description of the localized
semi-core wave functions.  In many applications like molecular
structure calculation semi-core pseudopotentials yield converged
results with comparable small basis sets even if the calculated total
energy is still far from its converged value. Therefore the inclusion
of semi-core electrons not inevitably requires the use of extremely
large basis sets.  In fact in our molecular calculations the highest
plane wave energy cutoff were needed for calculations with the
fluorine pseudopotential which has no semi-core electrons at all.

In many cases it is not quite clear, if semi-core electrons play an
important role or not. For most applications the need to use semi-core
pseudopotentials depends on the required accuracy and necessary
computational effort and should be tested carefully. Therefore we
constructed both type of pseudopotentials for most elements where
semi-core electrons can play a significant role in electronic
structure calculations.

\section{Molecules}

We tested our pseudopotentials by calculating the bond lengths of a
large number of molecules. In all calculations we used our scalar
relativistic pseudopotentials neglecting the terms for spin orbit
interaction. Whenever possible we tried to determine values for the
bond lengths, that are converged to $\approx$ 10$^{-3}$ Bohr. To
obtain this high level of accuracy extremely large boxes and high
plane wave energy cutoffs were needed so that for some molecules the
accuracy of the calculations was limited by our computational
resources.  The calculated bond lengths together with their
experimental values are listed in table~\ref{table2}. As a reference
for the quasi-exact LDA value we also list the bond lengths calculated
with {\small GAUSSIAN 94}\cite{g94} using a 6-311G++(3df,3pd) basis set (for the
$3d$-elements no $f$-polarization functions have been used). With a few
exceptions the values calculated with {\small GAUSSIAN 94} agree within
a few thousands of a Bohr with the LDA results published by Dickson
and Becke\cite{x4} and therefore should be close to the LDA limit.
For some molecules where no high precision basis sets are available we
took the all-electron results from Dickson and Becke. To estimate the
error arising from the pseudopotential approximation our calculated
values should be compared with these LDA results rather than with the
experimental bond lengths.  Unfortunately exact LDA values for
molecules containing heavier elements often are not available because
of the lack of a sufficiently accurate basis set.

The bond lengths calculated with our pseudopotentials, including
semi-core electrons where necessary, agree very well with the
all-electron values obtained with {\small GAUSSIAN 94} . The error of
the pseudopotential approximation for first row atoms is nearly ten
times smaller than the LDA error and for the heavier elements at least
comparable to the LDA error. In all cases except for the non semi-core
pseudopotentials the accuracy relative to the exact LDA value is,
however, better than the results obtained with standard Gaussian
6-31G$^*$ basis sets and it is comparable or better than the results
obtained with other all-electron methods. It must be mentioned that
our results especially for molecules with heavier elements are not
exactly comparable to the values obtained with {\small GAUSSIAN 94} or
the values of Dickson and Becke as our pseudopotentials also include
relativistic effects.

For some non semi-core pseudopotentials the error in the calculated
bond lengths is quite large. Nevertheless these pseudopotentials may
still be of interest for electronic structure calculations if no high
precision is required or the computational resources are
limited. These pseudopotentials require only small basis sets which is
sometimes a necessity for the study of large systems.

Our calculated bond lengths containing only first or second row atoms
also agree to within one or two thousands of a Bohr to those obtained
with nonrelativistic versions of these pseudopotentials that have
already been published\cite{x1} (differences in the case of HCN are
due to the choice of a different simulation box). This clearly
demonstrates that relativistic effects do not influence the bond
lengths for these molecules on a relevant scale.

\section{The Parameters}
In the following we list the parameters for all elements up to Rn.
The entries in Table I  have the following meaning:
 $$\begin{array}{lllllll}
\mbox{Element}&Z_{\rm ion}&r_{\rm loc}&C_1      &C_2      & C_3     & C_4 \\
       &       &r_0        &h^0_{1,1}&h^0_{2,2}&h^0_{3,3}&\\
       &       &r_1        &h^1_{1,1}&h^1_{2,2}&h^1_{3,3}&\\
       &       &           &k^1_{1,1}&k^1_{2,2}&k^1_{3,3}&\\
       &       &r_2        &h^2_{1,1}&h^2_{2,2}&h^2_{3,3}&\\
       &       &           &k^2_{1,1}&k^2_{2,2}&k^2_{3,3}&\\
       &       &\vdots     &           &           &           & \; .\\
\end{array}$$
Only the nonzero parameters are shown in table~\ref{table1}.
Parameters for elements marked with $^{sc}$ correspond to semi-core
pseudopotentials. In order to keep the table as small as possible the
coefficients $h^l_{i,j}$ and $k^l_{i,j}$ of the nonlocal projectors
for $i \not= j$ are not listed. To get the full parameter set the
missing $h^l_{i,j}$ and $k^l_{i,j}$ have to be calculated from the
$h^l_{i,i}$ and $k^l_{i,i}$. The relevant equations for the
$h^l_{i,j}$ are:
\begin{eqnarray}
\label{ort1}h^0_{1,2} &=& -\frac{1}{2}  \sqrt{\frac{3} {5}} h^0_{2,2}  \; ,\\
            h^0_{1,3} &=&  \frac{1}{2}  \sqrt{\frac{5}{21}}   h^0_{3,3} \; ,\\
            h^0_{2,3} &=& -\frac{1}{2}  \sqrt{\frac{100}{63}} h^0_{3,3} \; ,\\
            h^1_{1,2} &=& -\frac{1}{2}  \sqrt{\frac{5}{7}}     h^1_{2,2} \; ,\\
            h^1_{1,3} &=&  \frac{1}{6}  \sqrt{\frac{35}{11}}   h^1_{3,3} \; ,\\
            h^1_{2,3} &=& -\frac{1}{6}  \frac{14}{\sqrt{11}}   h^1_{3,3} \; ,\\
            h^2_{1,2} &=& -\frac{1}{2}  \sqrt{\frac{7}{9}}     h^2_{2,2} \; ,\\
            h^2_{1,3} &=&  \frac{1}{2}  \sqrt{\frac{63}{143}}   h^2_{3,3} \; ,\\
            h^2_{2,3} &=& -\frac{1}{2}  \frac{18}{\sqrt{143}}h^2_{3,3},\\
\label{ort2} h^l_{i,j} &=& h^l_{j,i}\; .
\end{eqnarray}
By this procedure one obtains a set of projector functions
\begin{equation}
P^l_k(r)= \sum_{i=1}^k \sum_{j=1}^k h^l_{i,j}p^l_{i,j}(r),\quad k=1,2,3
\end{equation}
that satisfy the orthogonality relation 
\begin{equation}
\int_0^\infty P_i^l(r) P_j^l(r)r^2dr=0 \quad\mbox{for}\quad i\not=j \; .
\end{equation}
Replacing $h^l_{i,j}$ by $k^l_{i,j}$ in eq.(\ref{ort1}-\ref{ort2})
gives the equations for the $k^l_{i,j}$.  We found that the
orthogonalization of the projectors improves the fitting procedure of
our pseudopotential rather than keeping all $h^l_{i,j}$ zero for
$i\not=j$ as was done in the former work of Goedecker {\sl et
al.}\cite{x1} Treating the $h^l_{i,j}$ as independent pseudopotential
parameters does not improve the results.

\section{Summary}
We developed a complete set of relativistic LDA pseudopotentials for
the whole periodic system up to Rn. The pseudopotentials are easy to
use as only a few parameters are necessary. All terms for both Fourier
and real space are given analytically and no tabulated functions are
needed. The pseudopotentials are highly accurate and transferable and
have been tested in extensive atomic and molecular calculations.

Gaussian type pseudopotentials for other exchange correlation
functionals or gradient corrected functionals can easily be
constructed using our LDA parameter sets as an initial guess. The
necessary programs are available from the authors upon request.

\section*{ACKNOWLEDGMENTS}
Financial support by the Hoechst AG is gratefully acknowledged.  We
thank Sverre Froyen and Mike Teter for making available their
relativistic atomic codes.

\clearpage
{\begin{table} 
 \caption{LDA pseudopotential parameters. The meaning of the entries
  is given in the text.\label{table1}}

\begin{tabular}{rrrrrrr}  \\ 

H  &   1         &    .200000 &  -4.180237 &    .725075 &&  \\
He &   2         &    .200000 &  -9.112023 &   1.698368 &&  \\

Li &   1         &    .787553 &  -1.892612 &    .286060 &&  \\
   &&    .666375 &   1.858811 &&&  \\  
   &&   1.079306 &   -.005895 &&&  \\  
   &&            &    .000019 &&&  \\  
Li$^{sc}$ &   3         &    .400000 & -14.034868 &   9.553476 &  -1.766488 &    .084370 \\
\end{tabular} \end{table}}

\begin{table} 
  \caption{Comparison of bond lengths of molecules calculated with our
pseudopotentials (PSP) and obtained with {\small GAUSSIAN 94} or
{\small NUMOL}\cite{x4} (AE) and the experimental data.  All {\small
GAUSSIAN 94} calculations were done with a 6-311++G(3df,3pd) basis set
if not otherwise mentioned. Bond lengths are given in
Bohr. \label{table2}}
\small
\begin{tabular}{llddc} 
\multicolumn{1}{c}{Molecule}&
\multicolumn{1}{c}{PSP}&
\multicolumn{1}{c}{AE}&
\multicolumn{1}{r}{Expt.}&
\multicolumn{1}{r}{Ref.} \\ \tableline
H$_2$           & 1.447               & 1.447     & 1.401 &\onlinecite{HH}  \\  
Li$_2$          & 5.100$^{sc}$, 5.016 & 5.120     & 5.051 &\onlinecite{HH} \\ 
LiH             & 3.029$^{sc}$, 2.929 & 3.029     & 3.015 &\onlinecite{HH} \\ 
Be$_2$          & 4.516$^{sc}$, 4.497 & 4.510     & 4.63  &\onlinecite{br40} \\ 
BF$_3$          & 2.477               & 2.469     & 2.470 &\onlinecite{LBII/7}\\  
BH              & 2.363               & 2.371     & 2.373 &\onlinecite{HH}  \\ 
CH$_4$          & 2.072               & 2.072     & 2.052 &\onlinecite{br43} \\
\end{tabular} 
$^a$:  {\small GAUSSIAN 94} with 3-21G basis set\\
$^b$:  {\small GAUSSIAN 94} with 6-311++G(3d,3pd) basis set\\
$^c$: calculated with {\small NUMOL} \\
\end{table}

\begin{table} 
  \caption{Transferability errors \label{table3}}
\begin{tabular}{lcccc}
Element&  state           & $\Delta E$ (Ha) & Error (Ha)               \\ \tableline
C
       & $2s^1 2p^3 3d^0$ & 0.3073                  & -0.40$\times$10$^{-3}$\\
       & $2s^2 2p^3 3d^0$ & 0.4294$\times$10$^{-1}$ &  0.11$\times$10$^{-3}$ \\
       & $2s^2 2p^1 3d^0$ & 0.3612                  & -0.44$\times$10$^{-3}$ \\
       & $2s^2 2p^2 3d^1$ & 0.3882                  &  0.95$\times$10$^{-4}$\\
Si
       & $3s^1 3p^3 3d^0$ &  0.2529                 &-0.20$\times$10$^{-3}$ \\
       & $3s^2 3p^3 3d^0$ & -0.5293$\times$10$^{-2}$ &   0.38$\times$10$^{-4}$\\
       & $3s^2 3p^1 3d^0$ &  0.2673     &  -0.28$\times$10$^{-3}$\\
\end{tabular}
\end{table}

\begin{figure}
\caption{Relativistic all-electron (solid) and pseudo (dashed) wave
functions of the valence electrons of gold. The difference between them
is shown by the dotted line on a logarithmic scale.}
\label{fig1}
\end{figure}

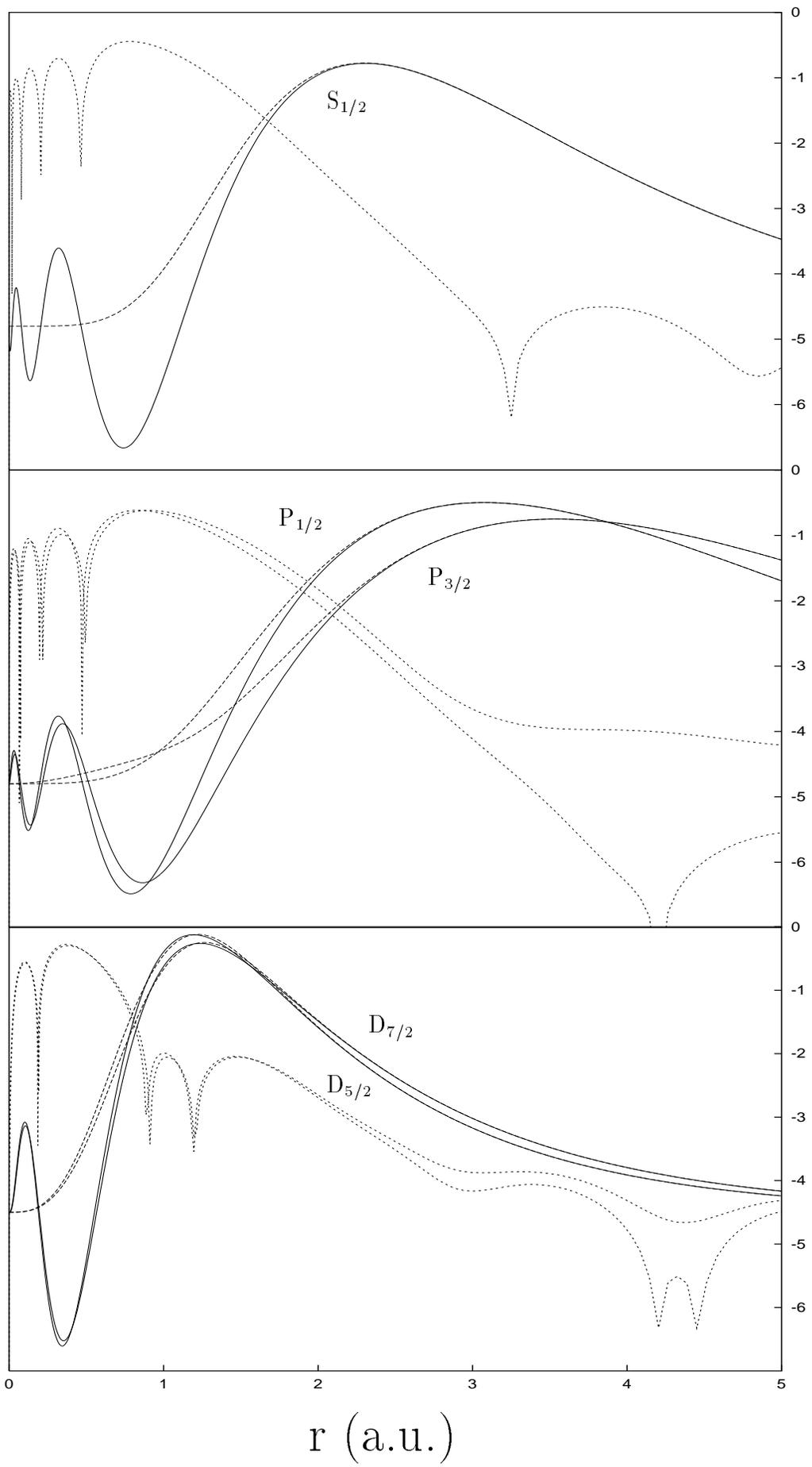
\end{document}